\documentclass[10pt]{article}

\usepackage{graphics} % for pdf, bitmapped graphics files
\usepackage{epsfig} % for postscript graphics files
\usepackage{times} % assumes new font selection scheme installed
\usepackage{amsmath} % assumes amsmath package installed
\usepackage{amssymb}  % assumes amsmath package installed
\usepackage{color}
\usepackage{cite}
\usepackage{algorithm}
\usepackage{algorithmicx}
\usepackage{algpseudocode}
\usepackage{booktabs}
\usepackage{url}
\usepackage{tikz}
\usepackage{neuralnetwork}
\usepackage{authblk}           % Author affiliations
\usetikzlibrary{shapes.geometric, arrows}
\usepackage{hyperref}
\usepackage{subcaption}

\usepackage{graphicx}
\usepackage{amsmath} % assumes amsmath package installed
\usepackage{amssymb}  % assumes amsmath package installed
\usepackage{subcaption}
\usepackage{url}
\usepackage{enumitem}
\usepackage{color}
\usepackage{algorithm}
\usepackage{algorithmicx}
\usepackage{algpseudocode}

\newcommand{\nx}{n_x}
\newcommand{\nz}{n_z}
\newcommand{\parvec}{z}
\newcommand{\rr}{\mathbb{R}}

\newcommand{\R}{\mathbb{R}}
\newcommand{\emb}{\mathrm{D}}
\newcommand{\AuE}{\mathcal{AE}_{\phi}}
\newcommand{\enc}{\mathrm{E}_{\phi_{e}}}
\newcommand{\dec}{\mathrm{D}_{\phi_{d}}}

\newtheorem{lemma}{Lemma}

\newtheorem{remark}{Remark}
\newtheorem{theorem}{Theorem}
\newtheorem{property}{Property}

\newtheorem{statement}{Problem Statement}

\newcommand{\Texp}{T_{\mathrm{exp}}}
\newcommand{\Tstop}{T_{\mathrm{stop}}}
\newcommand{\TMPC}{T_{\mathrm{s}}^{\mathrm{MPC}}}
\newcommand{\TCALCMPC}{T_{\mathrm{calc}}^{\mathrm{MPC}}}

\title{\LARGE \bf
Learning Low-Dimensional Embeddings for Black-Box Optimization
}

\author[1]{Riccardo Busetto}
\author[1]{Manas Mejari}
\author[1]{Marco Forgione}
\author[2]{Alberto Bemporad}
\author[1]{Dario Piga} % <-this % stops a space
\affil[1]{IDSIA Dalle Molle Institute for Artificial
Intelligence, SUPSI, Lugano-Viganello, CH 
        {\tt\small {name}.{surname}@supsi.ch}}%
\affil[2]{IMT School for Advanced Studies, Lucca, IT
        {\tt\small alberto.bemporad@imtlucca.it}}%
\date{} % Leave empty for no date, or use \date{\today}

\begin{document}

\maketitle

\begin{abstract}
Black-box optimization (BBO) is a valuable alternative to gradient-based methods when computing
derivatives of the objective function is not feasible. However, BBO often struggles with high-dimensional decision spaces and limited evaluation budgets. In this work, we propose a meta-learning approach that pre-computes a reduced-dimensional manifold capturing the optimal solutions of a class of optimization problems. For new problem instances drawn from the same class, BBO is then performed in the latent space, substantially reducing the computational effort required to reach near-optimal solutions. We further establish probabilistic upper bounds on the performance gap between solving the original problem and its reduced-dimensional counterpart. The effectiveness of the method is demonstrated both on benchmark optimization problems and on the automatic tuning of model predictive controllers.
\end{abstract}

\section{Introduction}

Black-box optimization (BBO)  aims to find the solution of an optimization problem where the objective function is unknown or lacks an explicit mathematical formulation. Instead, the function can only be evaluated through  queries, such as physical experiments or  simulations through complex computational models. However, in many real-world scenarios, these evaluations are expensive, noisy, or time-consuming. Therefore, a key goal of black-box optimization is to find near-optimal solutions while limiting the number of costly function evaluations.    
Due to these challenges, BBO relies on sample-efficient strategies that select query points to balance \emph{exploration} (searching unexplored regions) and \emph{exploitation} (refining promising solutions). Various techniques exist for tackling these problems, including  Kriging-based methods \cite{jeong2005efficient}, Bayesian Optimization (BO)~\cite{Shahriari16},  Mesh adaptive direct search \cite{audet2006mesh}, and recently proposed algorithms such as Set-Membership (SM) approaches~\cite{sabug2022smgo} and global  optimization  using inverse distance weighting and surrogate radial basis functions (GLIS)~\cite{Bem20}.

The primary limitation of the BBO algorithms mentioned above  is their  scalability with respect to the number of decision variables, restricting their applicability to small- to medium-size optimization problems.  
Black-box optimization in high-dimensional spaces still remains an open challenge, as the number of function evaluations required to cover the domain of decision variables increases exponentially with the number of dimensions. % 
In recent years, there have been a few works to mitigate this {curse of dimensionality} issue, aiming at developing methods that enhance the scalability of BBO in high dimensions by exploiting different assumptions 
about the structure of the underlying objective function. The seminal work \emph{Random EMbedding BO} (REMBO) \cite{REMBO16}  laid the foundation for addressing Bayesian Optimization in high-dimensional spaces. REMBO leverages the low effective dimensionality of the objective function and employs a random projection matrix to map the decision variables into a lower-dimensional subspace, where execution of BO is feasible. However, it has been shown that this approach may perform poorly even on some synthetic problems with a true low-dimensional linear subspace due to distortions introduced by the random linear projections~\cite{letham20}. 
Alternatively, the methods in~\cite{addGP15} and  \cite{rolland18additive} assume an additive structure of the objective function and decompose it into a sum of low-dimensional components, each depending on a small subset of variables. In \cite{ziomek23a}, data-independent rules are explored to identify effective additive decompositions.  In~\cite{RoFoPi20}, Bayesian optimization is implemented  iteratively by optimizing a subset of parameters while keeping the others fixed. The methodology is applied to tune the PID gains of a torque-controlled, 7-degree-of-freedom robot manipulator.

By leveraging the meta-learning paradigm~\cite{schmidhuber1987evolutionary}, our paper focuses on reducing the dimension of the decision space where BBO is performed. Meta-learning, or ``learning to learn'', involves acquiring knowledge from a class of related tasks to improve the learning process for new tasks within the same class. Following this principle, the key idea behind our approach is to learn a low-dimensional manifold where the optimal solutions of a class of similar optimization problems lie.  To identify this manifold, we learn an embedding that maps the optimal decision variables of a set of problems within the considered class into a lower-dimensional latent space. This embedding is obtained using an \emph{autoencoder} (AE) neural network, trained to minimize the reconstruction error between the optimal solutions in the original higher-dimensional space and the decoder output. The latent space of the trained AE thus serves as the desired low-dimensional representation.

An advantage of our approach is that the optimal solutions in the original input domain  can be precomputed offline using, for example, simulators with varying parameters and  computationally expensive optimization algorithms, e.g., evolutionary algorithms. This precomputed knowledge is then leveraged to solve new optimization problems by performing black-box optimization directly in the learned embedded space rather than the original higher-dimensional space. For instance, in tuning of Model Predictive Control (MPC)  hyperparameters, one can first compute optimal hyperparameters  using simulators of a class of parameterized systems. Then, when applying the controller to a real system, calibration  is performed efficiently through experiments using BBO algorithms such as Bayesian Optimization  or GLIS, searching over the precomputed embedded space rather than the original higher-dimensional domain. Although our methodology can, in principle, be applied in conjunction with any BBO method, in this paper we primarily focus on the GLIS algorithm.  The problem addressed in the paper is further motivated by considering a multiparametric programming problem~\cite{gal2010postoptimal} as a problem-class. Under suitable assumptions, the optimizer can be written explicitly as a function of the parameter vector, thereby defining a manifold where all the optimizers lie. 
 
As a further contribution, in this work we provide a theoretical result that establishes a probabilistic, distribution-free bound on the performance loss incurred when solving an optimization problem in the latent space instead of the original domain.  The result does not depend on the specific architecture used to construct the embedding, and can in principle be applied to neural autoencoders, random projections, or other dimensionality-reduction techniques. Furthermore, the bound does not compare the (generally unattainable) global optima of the two problems, but rather the solutions actually returned by the numerical algorithms employed in each space.
This bound is independent of the particular numerical optimization algorithms employed in the two spaces. For example, optimization in the original domain could be carried out with gradient-based methods or Particle Swarm Optimization, while the reduced problem may be solved using GLIS or other black-box algorithms.

The rest of the paper is organized as follows. 
In Section~\ref{sec:problem}, we introduce the class of black-box optimization problems considered in this work 
and formalize their reduced-dimensional counterparts. 
Section~\ref{sec:learningLatent} presents the proposed methodology for learning low-dimensional embeddings 
via autoencoders trained on meta-datasets of near-optimal solutions. 
Section~\ref{sec:meta-glis} describes the Meta-GLIS algorithm, which applies the GLIS method 
in the learned latent space. 
In Section~\ref{sec:sub-opt}, we provide theoretical results quantifying the sub-optimality gap 
between optimization in the latent and original domains.  
Section~\ref{sec:examples} reports numerical results obtained on a nonconvex optimization benchmark
and for the automated calibration of MPC hyperparameters.  Finally, Section~\ref{sec:conclusions} draws conclusions and discusses directions for future work.

\section{Problem description} \label{sec:problem}
We consider the following \emph{class} of optimization problems parameterized by $\theta$
\begin{equation}
    \label{eq:class_problems_x}
    f^\star_x(\theta) = \min_{x} f(x;\theta) \qquad
    \text{s.t.} \quad x \in \mathcal{X} 
\end{equation}
where, for a given parameter  $\theta \in \mathbb{R}^{n_\theta}$, $f(x;\theta): \mathbb{R}^{\nx} \to \mathbb{R}$ is a function of the optimization variable $x \in \mathcal{X} \subseteq \mathbb{R}^{\nx}$.  The parameter $\theta$ represents a set of conditions that affect the optimization problem. These conditions may correspond to simulation settings, system configurations, or operational scenarios in real-world applications, which in turn influence the structure of the function $f(x; \theta)$.  

We assume that  an analytical expression of $f(x;\theta)$ is not available, but the function can be evaluated at any feasible query input $x \in \mathcal{X}$ and for any parameter $\theta$. This may be the case, for instance, when $\theta$ describes different configurations of the environment and  $x$ is a design variable to be computed in order to optimize a metric $f$, which in turn can be evaluated only through the execution of a real experiment or high-fidelity simulation.  
The set $\mathcal{X}$ represents known constraints on $x$.  
For the theoretical analysis of our approach, we will denote by $p(\theta)$ the distribution from which the values of the parameter $\theta$ are sampled (e.g., a uniform distribution over a hyperbox of parameters). For the practical implementation, as clarified in Section~\ref{sec:methodology}, we only assume access to a finite set of samples drawn from $p(\theta)$.

\begin{statement}
Given the class of optimization problems~\eqref{eq:class_problems_x} in $n_x$ variables
and a desired reduced dimension $\nz < \nx$, \emph{learn} a  map $\emb: \mathbb{R}^{\nz} \to \mathbb{R}^{\nx}$ such that the optimization of $f$ can be carried out effectively with respect to the lower-dimensional embedding $z\in\mathbb{R}^{\nz}$:
\begin{equation}
   \label{eq:class_problems_z}
     z^\star(\theta)  \in \arg\min_{z\in\mathcal{Z}} f(\emb(z);\theta),\\
\end{equation}
with 
\[
    f(\emb(z^{\star}(\theta));\theta)\approx f^\star_x(\theta),\quad \forall\theta\in\mathbb{R}^{n_\theta},
\]

where the feasible set    
$\mathcal{Z} = \{z \in \R^{n_z}: \emb(z) \in \mathcal{X}\}$.
\end{statement}
In other words, our goal is to find a map $\emb$ such that we can retrieve an approximation $\emb(z^{\star}(\theta))$ of an optimizer $x^{\star}(\theta)$ of the original problem \eqref{eq:class_problems_x} from the solution $z^{\star}(\theta)$ of the simpler problem~\eqref{eq:class_problems_z}.

\section{Methodology}
\label{sec:methodology}

\subsection{Meta-dataset creation}
We generate a set of $N$  parameters $ \{\theta^{(i)}\}_{i=1}^{N}$ by sampling from the  parameter distribution, \emph{i.e.},  $\theta^{(i)} \sim p(\theta)$. For each $\theta^{(i)}$, we solve \eqref{eq:class_problems_x} by using any sample-based  optimization algorithm (such evolutionary learning, MCMC with simulated annealing, Bayesian Optimization, GLIS)  and we gather a set $\{x^{(i)}_k, f^{(i)}_k\}_{k=1}^{K}$of $K$   {\it near-optimal} vectors $x^{(i)}_k$,  and the corresponding  function values  $f^{(i)}_k = f(x^{(i)}_k, \theta^{(i)})$, with $x^{(i)}_l \neq x^{(i)}_m, \forall l\neq m$, $l,m =1, \ldots, K$. 
For a given $\theta^{(i)}$, we refer to as near-optimal vectors as the best  $K$ candidate points explored during the iterations of the numerical algorithms used to solve \eqref{eq:class_problems_x}. 
We denote the meta-dataset as
\vspace{-0.0cm}
\begin{equation}
\label{eq:meta_dataset}
\mathcal{D} := \left\{\theta^{(i)}, \{x^{(i)}_k, f^{(i)}_k\}_{k=1}^{K} \right\}_{i=1}^{N}.
\end{equation}

\subsection{Learning a latent representation via autoencoders} \label{sec:learningLatent}
An \emph{autoencoder} (AE)~\cite{hinton06}  is trained to map an input vector $x \in \R^{\nx}$ to itself, while passing through a \emph{bottleneck} layer of size  $\nz < \nx$.  
Let us consider an autoencoder $\AuE: \R^{\nx} \to \R^{\nx}$ with tunable parameters $\phi$, which is composed of an encoder $\enc: \R^{\nx} \to \R^{\nz}$  mapping $x$ to a lower dimensional latent space vector $z$ and a decoder $\dec: \R^{\nz} \to \R^{\nx}$ mapping  the latent vector $z$ to reconstruct the original vector $x$.
Given the meta-dataset $\mathcal{D}$ in~\eqref{eq:meta_dataset}, we train the autoencoder  $\AuE$  by minimizing the following weighted  reconstruction error over the parameters $\phi = \{\phi_e, \phi_d\}$:
%\vspace{-.45cm}
\begin{align}\label{eq:AE_loss_sample}
\mathcal{L}(\phi) = \min_{\phi} \frac{1}{N}\sum_{i=1}^{N}  \sum_{k=1}^{K}w^{(i)}_k\|x^{(i)}_k - \AuE(x^{(i)}_k) \|^2_2,
\end{align}
where the weights $w^{(i)}_k$ are chosen to give relatively more importance to the best candidate points $x_k^{(i)}$. In particular, given a hyperparameter $\lambda \in [0, 1)$, we define the weights as follows:
%\vspace{-.1cm}
\[
\begin{aligned}
    (j_1^{(i)},\ldots,j^{(i)}_K) & := \mathrm{argsort}\left(\{f_k^{(i)}\}_{k=1}^K\right),\\
    w^{(i)}_k &:= \lambda^{j_k^{(i)}},\quad k=1,\ldots,K, 
\end{aligned}
\]
% \vspace{-.2cm}
where the function $\mathrm{argsort}(\cdot)$ sorts the argument values in ascending order.  

After training the autoencoder, black-box optimization is performed over $z$ in a lower-dimensional space. The decoder $\emb_{\phi_d}$ then maps the latent variable $z$ back to the original high-dimensional space, where the function $f(x;\theta)$ is evaluated.  In the following paragraph, we provide more details on the application of the proposed approach for the implementation of GLIS in a low-dimensional space. Nevertheless, the same ideas can also be applied to other BBO algorithms, such as Bayesian Optimization.

\begin{remark} \label{remark:bounds_outputs}
In order to ensure that the output of the autoencoder $\AuE$ lies within the feasible set $\mathcal{X}$, penalty terms can be added to the loss function in \eqref{eq:AE_loss_sample}. Alternatively, the last layer of the decoder $\dec$ can be designed to enforce constraint satisfaction. For instance, in the case of box constraints, scaled and shifted sigmoid activation functions can be used in the last layer of the decoder to enforce the specified bounds.
\end{remark}

\begin{remark}
Similarly to the discussion in Remark \ref{remark:bounds_outputs}, bounds on the latent variables can also be enforced. In particular, a sigmoid activation function is used in the last layer of the encoder to rescale the latent variable $z$ such that it belongs to the box $\mathcal{Z}=[0, \ 1]^{n_z}$. This restricts the search space over the latent variable domain, as discussed in the next section.
\end{remark}

\subsection{Meta-GLIS: GLIS over  a low-dimensional space} \label{sec:meta-glis}

In this section, we describe how to apply the proposed methodology to implement \emph{Meta-GLIS}, a lightweight version of GLIS~\cite{Bem20} based on meta learning that operates in a low-dimensional latent space.  
First, we recall that the  encoder $\enc$ and decoder $\dec$ are pretrained (e.g., in a simulation environment) using the meta-dataset $\mathcal{D}$ defined in \eqref{eq:meta_dataset}.  
Given a query parameter $\bar{\theta}$ (which represents, for instance,  the  conditions of a real-world problem, rather than a simulated one), the goal is to solve the original optimization problem \eqref{eq:class_problems_x} for $\theta = \bar{\theta}$. Outside the simulation environment, evaluating the objective function $f(x, \bar{\theta})$ is assumed to be costly. Therefore, Meta-GLIS is expected to provide   a  near-optimal solution within a  limited number of function evaluations.

The Meta-GLIS algorithm, summarized in Algorithm \ref{algo:GLISlite}, follows a standard surrogate-based black-box optimization strategy. First, an initial set of samples of $M_\text{init}$ samples is drawn from the latent space domain $\mathcal{Z} = [0, \ 1]^{n_z}$ using Latin Hypercube Sampling (LHS)~\cite{MBC79} (Step 2). The function $f(\dec(z);\bar{\theta})$ is then evaluated at these initial points, forming the initial dataset $\mathcal{D}_{M}$ (Step 3).  In the iterative optimization loop (Step 4), a \emph{surrogate model} $\hat f: \mathcal{Z} \rightarrow \mathbb{R}$ is constructed to approximate the objective function based on the dataset $\mathcal{D}_{M} = \{z_{j}, f(\dec(z_{j});\bar \theta)\}_{j=1}^M$  (Step 4.1). As in the original formulation of GLIS,  a linear combination of Radial Basis Functions (RBFs)~\cite{Gut01} is used to create the surrogate function. Specifically, 
\begin{equation}
\hat f(\parvec)=\sum_{k=1}^{M}\beta_{j}\phi(\mu d(\parvec,\parvec_k)),
\label{eq:rbf}
\end{equation}
where $\phi:\rr\to\rr$ is an RBF, $d:\mathcal{Z}\times\mathcal{Z}\to\rr$ is the squared
Euclidean distance $
d(\parvec,\parvec_k)=\|\parvec-\parvec_k\|_2^2$, 
$\mu>0$ is a scalar hyper-parameter defining the shape of the RBF. The unknown coefficients  $\beta=[\beta_{1}\ \ldots\ \beta_{M}]^{T}$ are  computed 
by fitting the surrogate $\hat f(\parvec)$ to the dataset $\mathcal{D}_M$ through the minimization  of the  regularized squared error: 
	\begin{align} \label{eqn: reg_LS}
		\sum_{j=1}^{M}\Big\| f(\dec(z_{j};\bar \theta)) - \underset{\hat f(\parvec_j)}{\underbrace{\sum_{k=1}^{M}\beta_j\phi(\mu d(\parvec_j,\parvec_k))}} \Big\|^2 + \gamma  \left\|  \beta_j \right\|^2, 
	\end{align} 
where the quadratic regularization term (weighted by the  hyperparameter $\gamma>0$) is added to guarantee strict convexity of~\eqref{eqn: reg_LS}. 
Some RBFs commonly used are $\phi(\mu d)=\frac{1}{1+(\mu d)^2}$
(\emph{inverse quadratic}) and $\phi(\mu d)=e^{-(\mu d)^2}$ (\emph{squared exponential kernel}). Both the   hyperparameter $\mu$ and $\gamma$ are tuned through cross-validation.

The \textit{acquisition function} $a(z)$ is then constructed (Step 4.2) as:  
\begin{align} \label{eq:ac}
    a(z) = \hat{f}(z) - \delta h(z),
\end{align}  
where $h(z)$ is an exploration-promoting function, and $\delta > 0$ is a tuning hyperparameter that balances exploitation and exploration in the acquisition function. According to GLIS \cite[Section 4]{Bem20}, the exploration term is defined by the following \emph{Inverse Distance Weighting} (IDW) function
\begin{equation}
    h(z) =
    \begin{cases}
        0 & \text{if } z \in \mathcal{D}_M, \\
        \frac{2}{\pi}\tan^{-1} \left( \frac{1}{\sum_{i=1}^{M} \frac{1}{\|z - z_i\|^2}} \right) & \text{otherwise}.
    \end{cases}
\end{equation}
The acquisition function~\eqref{eq:ac} is then minimized to determine the next sampling point $z_{M+1}$ and the corresponding decoded value $x_{M+1} = \dec(z_{M+1})$, at which the objective function is evaluated (Steps 4.3 and 4.4). The process continues until the maximum number of iterations is reached. The best solution is then returned.

\begin{algorithm}[!ht]
\caption{\textbf{Meta-GLIS}}\label{algo:GLISlite}
\textbf{Input}: number $M_{\text{init}}$ of initial samples; maximum number $M_{\max} \geq M_{\text{init}}$
of function evaluations; pre-trained encoder $\enc$ and decoder $\dec$; query parameter $\bar{\theta}$.
\hrule\vspace{0.3em}

\begin{algorithmic}[1]
    \State Set $M \gets M_{\text{init}}$;
    \State Generate $M$ random initial samples $Z = [z_1, \dots, z_M]$ using LHS from $\mathcal{Z}$;
    \State Create dataset $\mathcal{D}_{M} \gets \{(z_{j}, f(\dec(z_{j};\bar{\theta})))\}_{j=1}^M$;
    \While{$M < M_{\max}$}:
        \begin{enumerate}[label=4.\arabic*,
                  leftmargin=1.5em,
                  labelsep=0.5em,
                  itemsep=0.2ex,
                  font=\small]
              \item Compute\,surrogate\,$\hat{f}(z)$\,of\,$f(\dec(z);\bar{\theta})$\,from\,$\mathcal{D}_{M}$;
              \item Compute acquisition function $a(z)$ \hfill (see Eq.~\eqref{eq:ac});
              \item Compute next input $z_{M+1} = \arg\min\limits_{z \in \mathcal{Z}} a(z)$;
              \item Evaluate $f(\dec(z_{M+1}); \bar{\theta})$;
              \item Augment the dataset:
                 $$\mathcal{D}_{M+1} \gets \mathcal{D}_{M} \cup \{(z_{M+1}, f(\dec(z_{M+1});\bar{\theta}))\}$$
              \item $M \gets M + 1$;
        \end{enumerate}
    \EndWhile;
    \State Compute best solution: 
    \[
    \begin{aligned}
       z^* &= \arg\min_{z \in \mathcal{D}_M} f(\dec(z);\bar{\theta}), \\
       x^* &\gets \dec(z^*),\ 
       f^* \gets f(x^*;\bar{\theta}).
    \end{aligned}
    \]
\end{algorithmic}
\hrule\vspace{0.3em}
\textbf{Output}: Best function value $f^*$ and corresponding input $x^*$.
\end{algorithm}

%%%%%%%%%%%%%%%

\section{Sub-optimality quantification} \label{sec:sub-opt}

The goal of  this section is to establish probabilistic theoretical bounds that quantify the gap between solving the original problem \eqref{eq:class_problems_x} and the reduced-dimensional  problem \eqref{eq:class_problems_z} over the latent set $\mathcal{Z}$.

\subsection{Definitions}

For a given parameter $\theta$ taking values in a measurable set $\Theta \subseteq \mathbb{R}^{n_{\theta}}$ 
and a random variable $\xi \in \Xi \subseteq \mathbb{R}^{n_\xi}$, 
let $\mathcal{A}_x : \Theta \times \Xi \to \mathcal{X}$ be a deterministic  map from $(\theta,\xi)$ to a value $x=\mathcal{A}_x(\theta,\xi)$, $x\in\mathcal{X}$. In our work, $\mathcal{A}_x$  represents  a numerical optimization algorithm (e.g., gradient descent, 
particle swarm optimization, Bayesian optimization),  $\xi$ the random realizations characterizing 
the algorithm (e.g., initial conditions, mini-batches in stochastic gradient descent, etc.), and 
$\mathcal{A}_x(\theta,\xi)$ 
the solution returned by  $\mathcal{A}_x$ when applied to problem~\eqref{eq:class_problems_x} 
with parameters  $\theta$ and for random algorithm's parameters $\xi$.   
 Alternatively, $\mathcal{A}_x$ can be the mapping that returns a global minimizer of 
problem~\eqref{eq:class_problems_x}. In this case,    $x^\star(\theta) = \mathcal{A}_x(\theta,\xi)$ 
depends only on the parameters $\theta$ characterizing the objective function $f$\footnote{
In case of multiple global optima, $x^\star(\theta)$ may also depend on some random variables $\xi$ if $x^\star(\theta)$ is selected at random from the set of global optimizers.}.  

Analogously, let $\mathcal{A}_z : \Theta \times \Omega \to \mathcal{Z}$ be a deterministic mapping such that $\mathcal{A}_z(\theta,\omega)$ represents the solution of the reduced-dimensional problem~\eqref{eq:class_problems_z} returned by algorithm $\mathcal{A}_z$,   with $\omega \in \Omega$ capturing the random settings involved in the reduced-space optimization procedure. 
Note that $\mathcal{A}_x$ and $\mathcal{A}_z$ do not need to be the same numerical algorithms.

We measure the relative gap in performance when optimizing in the reduced space via the \emph{relative performance gap} defined as:
\begin{align} \label{eqn_perc_gap}
\Psi(\theta, \xi,\omega) := 
\frac{f(\emb(\mathcal{A}_z(\theta,\omega));\theta) - f(\mathcal{A}_x(\theta,\xi);\theta)}
{|f(\mathcal{A}_x(\theta,\xi);\theta)| + \epsilon},
\end{align}
where $\epsilon > 0$ is a small constant introduced to avoid possible divisions by zero.  
In the special case
\begin{align}
\mathcal{A}_x(\theta,\xi) = x^\star(\theta), 
\qquad 
\mathcal{A}_z(\theta,\omega) = z^\star(\theta),
\end{align}
the performance gap $\Psi(\theta, \xi,\omega)$ measures the theoretical loss between the optimal solution of problem~\eqref{eq:class_problems_x} and that of the reduced-dimensional 
problem~\eqref{eq:class_problems_z}. However, this value can be computed only if problems~\eqref{eq:class_problems_x} and~\eqref{eq:class_problems_z} can be solved exactly at the global optimum.  More generally, when $\mathcal{A}_x$ and $\mathcal{A}_z$ denote numerical optimization algorithms, 
the performance gap may also take negative values as the global optimum of 
problem~\eqref{eq:class_problems_x} might  not be reached by  the algorithm $\mathcal{A}_x$.

Let 
$p(\xi)$ and $p(\omega)$ be  distributions over the random variables $\xi$ and $\omega$, and  $\{(\theta^{(i)},\allowbreak \xi^{(i)},\allowbreak \omega^{(i)})\}_{i=1}^m$ be a validation set of i.i.d.\ draws from 
 product measure $p(\theta)\otimes p(\xi)\otimes p(\omega)$.    
The validation set thus corresponds to $m$ realizations of the performance gap $\Psi$, namely, with $i=1,\dots,m$, 
\begin{align}
f^{(i)} := f(x(\theta^{(i)},\xi^{(i)});\theta^{(i)}), 
\quad  
\Psi^{(i)} := \Psi(\theta^{(i)},\xi^{(i)},\omega^{(i)}). 
%\quad i=1,\dots,m.
\end{align}

Finally, let $(\theta^{\rm new},\xi^{\rm new},\omega^{\rm new})$ be an independent draw from 
$p(\theta)\otimes p(\xi)\otimes p(\omega)$, not used during training or validation, 
with corresponding objective value $f^{\rm new}$ and performance gap $\Psi^{\rm new}$.

We denote by $\mathbb{P}$ the underlying probability measure induced by the product distribution 
$p(\theta)\otimes p(\xi)\otimes p(\omega)$. 
Accordingly, for any measurable set $A \subseteq \mathbb{R}$, we write 
\begin{align}
\mathbb{P}_\Psi(A) := \mathbb{P}(\Psi \in A) 
= \mathbb{P}\left(   \Psi^{-1}(A)\right)
\end{align}
to denote the probability distribution of the random variable $\Psi$.  Similarly, let $S=\{\Psi^{(i)}\}_{i=1}^m$ denote the validation set.  We write $\mathbb{P}_S$ 
for the probability distribution of $S$, induced by the i.i.d.\ draws 
$\{(\theta^{(i)},\xi^{(i)},\omega^{(i)})\}_{i=1}^m \sim p(\theta)\otimes p(\xi)\otimes p(\omega)$.

\subsection{Generalization bounds on the performance gap}

\begin{theorem}[Generalization bounds] \label{theorem:gap}

For any $\alpha,\delta \in (0,1)$, let
$
k^\star := \Big\lceil m\,(1-\alpha+\varepsilon_m)\Big\rceil$ and 
$\varepsilon_m := \sqrt{\tfrac{\log(2/\delta)}{2m}}$. 
Assume that the number of validation samples $m$ is large enough so that $\alpha \geq  \varepsilon_m$ (or equivalently,  $k^\star \le m$).  
Then, with probability at least $1-\delta$  over the validation set 
$S=\left\{\Psi^{(i)} \right\}_{i=1}^{m}$, we have
\begin{align} \label{eqn:Psi_probl}
\mathbb{P}_{\Psi}\!\left(\Psi \le \Psi_{(k^\star)}\right) \;\ge\; 1 - \alpha,
\end{align}
where  $\Psi_{(k)}$ denotes the $k$-th 
order statistic of the validation values, i.e., if 
\begin{align}
\Psi_{(1)} \le \Psi_{(2)} \le \dots \le \Psi_{(m)},
\end{align}
then $\Psi_{(k)}$ is the $k$-th smallest value in the ordered sequence.  
Equivalently,
\begin{align} \label{eqn:Psi|S_probl}
\mathbb{P}_S\Big( \, \mathbb{P}_{\Psi}\big(\Psi 
   \le \Psi_{(k^\star)} \,\big|\, S\big) \;\ge\; 1-\alpha \,\Big) \;\ge\; 1-\delta.
\end{align}
\hfill $\blacksquare$
\end{theorem}

The proof of the theorem  is reported in the Appendix.  \\

 Theorem~\ref{theorem:gap} has the following interpretation. 
To construct the validation set, we generate $m$ i.i.d.\ realizations 
$\{(\theta^{(i)},\xi^{(i)},\omega^{(i)})\}_{i=1}^m$ from the distribution product  
$p(\theta)\otimes p(\xi)\otimes p(\omega)$. 
For each triple $(\theta^{(i)},\xi^{(i)},\omega^{(i)})$, we run the two optimization algorithms 
$\mathcal{A}_x$ in the original space (e.g., PSO) and $\mathcal{A}_z$ in the reduced latent space 
(e.g., GLIS), and we compute the corresponding performance gap 
\begin{align}
\Psi^{(i)} := \Psi(\theta^{(i)},\xi^{(i)},\omega^{(i)}), 
\qquad i=1,\dots,m.
\end{align}
The theorem guarantees that, with probability at least $1-\delta$ over the draw of the validation set $S$, 
the order statistic $\Psi_{(k^\star)}$ 
provides an upper bound on the $(1-\alpha)$-quantile of the true distribution of $\Psi$. 
Equivalently, for a new independent draw 
$(\theta^{\rm new},\xi^{\rm new},\omega^{\rm new}) \sim p(\theta)\otimes p(\xi)\otimes p(\omega)$ and corresponding performance gap  $\Psi^{\rm new} = \Psi(\theta^{\rm new},\xi^{\rm new},\omega^{\rm new})$, 
we have
\begin{align}
\mathbb{P}_\Psi\!\left(\Psi^{\rm new} \;\le\; \Psi_{(k^\star)} \,\Big|\, S\right) \;\ge\; 1-\alpha,
\end{align}
and this condition  holds with probability at least $1-\delta$ 
with respect to the randomness of the validation set $S$.

\section{Numerical examples} \label{sec:examples}
We evaluate the performance of Meta-GLIS  on two numerical examples. The first involves the Rosenbrock function~\cite{Ros60}, a well-established benchmark problem in nonlinear optimization. The second focuses on the automated calibration of a set of MPC tuning knobs.  
Computations  are carried out on a server equipped with an NVIDIA RTX 3090 GPU. To guarantee reproducibility of the results, all Python scripts  are made available in the GitHub repository \url{https://github.com/buswayne/meta-glis}.  

\subsection{Rosenbrock function benchmark}
The  Rosenbrock function to be minimized  is defined as:
\begin{equation}\label{eq:rosenbrock}
    f(x) = \sum_{i=1}^{\nx-1} \left[ \theta_1 \left( x_{i+1} - x^2_i \right)^2 +
    \theta_2(\theta_{3,i} - x_i)^2\right],
\end{equation}
where $\theta_1 \in \mathbb{R}, \theta_2 \in \mathbb{R}, \theta_3 \in \mathbb{R}^{\nx-1}$ are random realization sampled, respectively, from the following uniform distributions $\mathcal{U}_{[10, 1000]},
\mathcal{U}_{[0.1, 10]}, (\mathcal{U}_{[0.1, 10]},\ldots,\mathcal{U}_{[0.1, 10]})$. For the nominal case, $\theta = \left[ 100, 1, \boldsymbol{1} \right]$, where $\boldsymbol{1} \in \mathbb{R}^{n_x}$ is a vector with all entries equal to 1, the function is non-convex and has a narrow, curved valley containing the global minimum at $(x_1^\star, \ldots, x_{\nx}^\star) = (1, \ldots, 1)$. %The steep walls and flat valley make it challenging for optimization algorithms that rely on gradient-based methods.

The meta-dataset is obtained from $N=500$  instances of the parameter $\theta$, sampled from the uniform distributions described above, with $\nx=20$. We solved each problem by employing a \emph{differential evolution} algorithm~\cite{storn1997differential} with 1000 generations over the feasible input domain $\mathcal{X}=[-2.5,2.5]^{20}$. 
%a population size of $10$, a number of generation of $1000$ and bounds $\mathcal{X}=[-2.5,2.5]^{20}$. 
We then created a meta-dataset $\mathcal{D} = \left\{\theta^{(i)}, \{x^{(i)}_k, f^{(i)}_k\}_{k=1}^{1000} \right\}_{i=1}^{500}$.
Similarly, we generated a separate test dataset $\mathcal{D}^{\mathrm{test}} = \left\{\theta^{(i)}, \{x^{(i)}_k, f^{(i)}_k\}_{k=1}^{1000} \right\}_{i=1}^{100}$ to evaluate the reconstruction capabilities of an autoencoder with latent dimension $\nz = 3$,  when trained as discussed in Section \ref{sec:learningLatent}. The encoder and decoder are fully connected neural networks, each with two hidden layers. The encoder has $n_x$ inputs, a first hidden layer with 128 units, a second hidden layer with 64 units, and $n_z$ outputs. The decoder has a symmetric structure. The time required to train the autoencoder is %$1229$ s (approximately 20.4 minutes).
approximately 20 minutes.
  
\begin{figure}[!tp]
    \centering
\includegraphics[width=\linewidth]{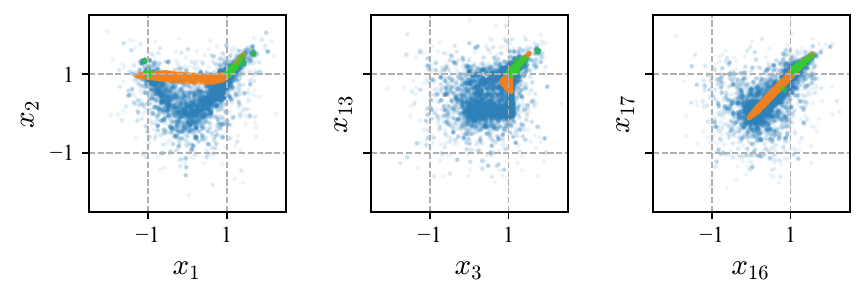}
    \vspace{-0.0cm}
    \caption{Learned manifold of the trained autoencoder  over test functions: candidate solutions generated by the differential evolution algorithm (blue points); optimal solution for each problem instance (green points); output of the autoencoder  for each candidate solution (orange points).    
   }
\label{fig:rosenbrock_autoencoder}
\end{figure}

%\vspace{-0.5cm} 
In Fig.~\ref{fig:rosenbrock_autoencoder} we can see the candidate solutions generated by the differential evolution algorithm (blue points), along with the optimal solution for each problem instance (green points) and the  autoencoder output for each candidate solution (orange points). We can see that the reconstructed inputs lie in a manifold with reduced dispersion, which accurately describes the region where optimal solutions lie.

The Meta-GLIS algorithm is applied in the learned $n_z=3$-dimensional latent space, with $M_{\rm init}=2n_z=6$ initial random points generated using LHS within the box $\mathcal{Z} = [0, \ 1]^{3}$. For comparison, optimization is also performed using GLIS over the original $20$-dimensional input domain. In this case, $2n_x=40$ initial points are randomly generated within the box $[-2.5, \ 2.5]^{20}$ using LHS. A Monte Carlo analysis  is conducted on 100 newly generated problem instances that were not used for training the autoencoder. 

\begin{figure}
    \centering
\includegraphics[width=\linewidth]{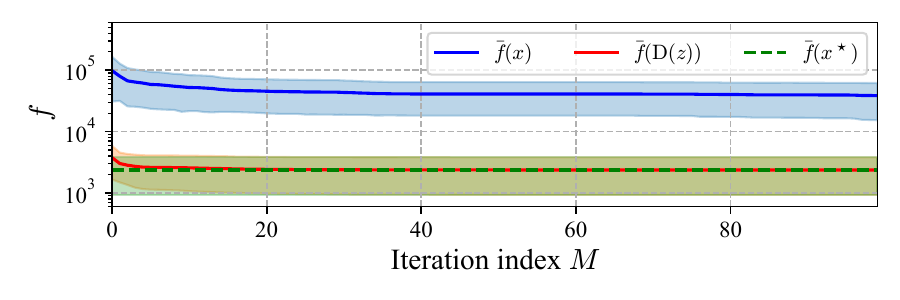}
    \vspace{-0.0cm}
    \caption{Comparison of the average performance obtained  with GLIS (blue) and Meta-GLIS (red) over $N^{\mathrm{test}} = 100$ runs. Global optimum (green).  Shaded areas correspond $\pm$ one standard deviation over $100$ Monte Carlo runs.}
    \label{fig:rosenbrock_iterations}
\end{figure}

The obtained numerical results are reported in Fig. \ref{fig:rosenbrock_iterations}, which shows the best value of the Rosenbrock objective function achieved at each iteration of the  GLIS and Meta-GLIS algorithm. The figure presents both the average and standard deviation over 100 Monte Carlo runs. The results highlight the advantages of Meta-GLIS, which achieves near-optimal solutions after approximately 20 iterations, whereas  GLIS remains  far from the global optimum even after 100 iterations. This underscores the well-known advantages of running  GLIS, and more in general BBO, in low-dimensional spaces, due to the reduced curse of dimensionality caused by exploration.

\begin{figure}
    \centering
    \includegraphics[width=\linewidth]{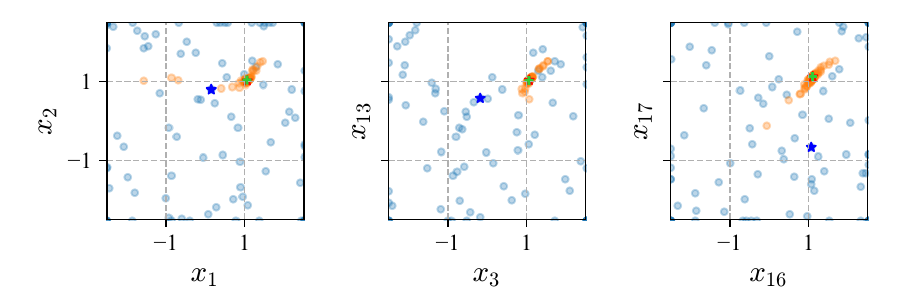}
    \caption{Sampled points by GLIS (blue circles) and Meta-GLIS (orange circles), along with corresponding best solutions (stars). The global optimal solution (green cross) overlaps with the orange star.}
    \label{fig:rosenbrock_glis}
\end{figure}

This advantage can be formally quantified by applying the theoretical guarantees of Theorem~\ref{theorem:gap}. 
We generated $M=1000$ validation instances of the optimization problem and solved them using PSO, GLIS, and Meta-GLIS, 
which allowed us to compute the corresponding probabilistic bounds on the performance gap $\Psi$.  

\begin{figure}[t]
    \centering
    
    % First subplot
    \begin{subfigure}{\linewidth}
        \centering
        \includegraphics[width=\linewidth, trim=0 0 0 1.5cm, clip]{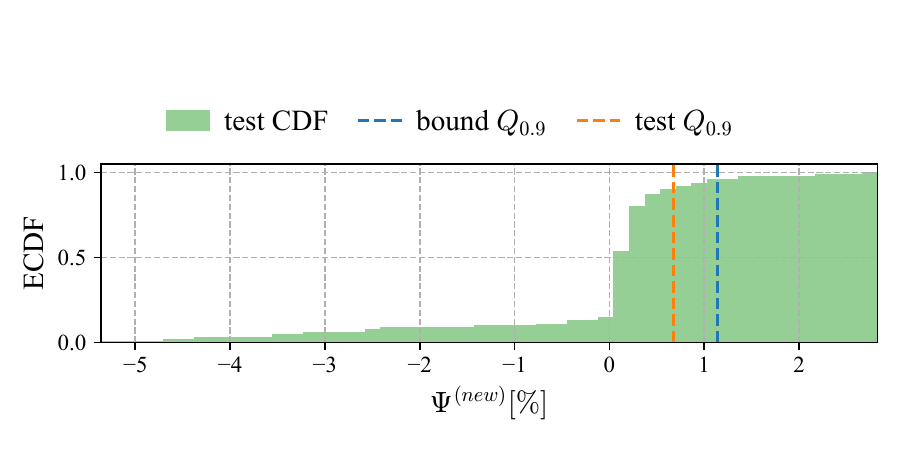}
        \caption{Meta-GLIS vs. PSO}
        \label{fig:cdf_pso}
    \end{subfigure}
    
    \vspace{0.5cm} % space between subplots
    
    % Second subplot
    \begin{subfigure}{\linewidth}
        \centering
        \includegraphics[width=\linewidth, trim=0 0 0 1.5cm, clip]{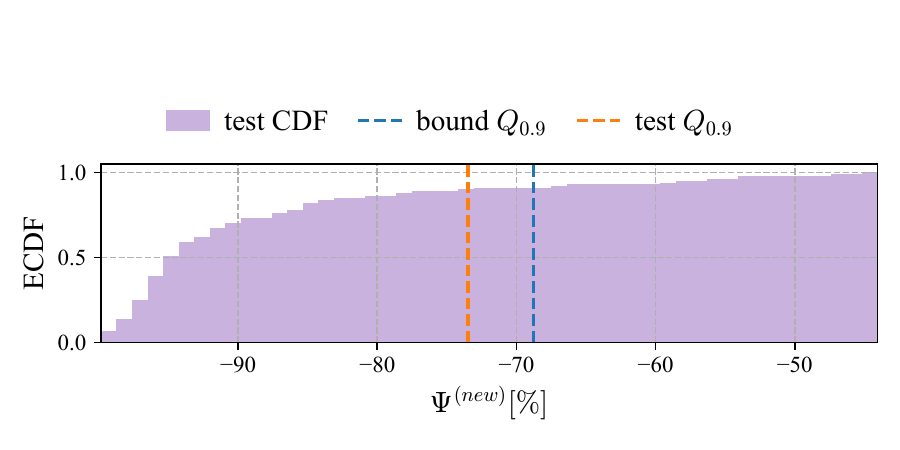}
        \caption{Meta-GLIS vs. GLIS}
        \label{fig:cdf_normal}
    \end{subfigure}
    
    \caption{Empirical CDFs of the performance gap using Meta-GLIS compared with  PSO (panel (a)) and GLIS (panel (b)) over $100$ test instances.  Theoretical bound from Theorem~\ref{theorem:gap} (blue line);  empirical 90\%-quantile on the test set (orange line).}
    \label{fig:cdf_bounds}
\end{figure}

For $\delta = 0.05$ and $\alpha = 0.1$, the computed bound on the performance gap between Meta-GLIS and PSO is $1.14\%$ (blue dashed line in Fig.~\ref{fig:cdf_pso}).
Thus, with probability at least $95\%$ (i.e., $1-\delta$), the likelihood that a new instance optimized with Meta-GLIS leads to a performance gap w.r.t. PSO lower than   $1.14\%$  is at least  $90\%$ (i.e., $1-\alpha$).
This result is consistent with the empirical $90$-th percentile on the test set (orange dashed line), which gives a smaller gap of $0.68\%$, thereby validating the bound. 

By applying the same reasoning to compare standard GLIS (operating on the full space $x$) against Meta-GLIS, 
we obtain the result shown in Fig.~\ref{fig:cdf_normal}. 
In this case, Meta-GLIS demonstrates a clear advantage: $90\%$ of the test instances achieve at least a 
 68.87\% relative improvement  when optimization is performed in the latent space.  This result can qualitatively be appreciated also in Fig.~\ref{fig:rosenbrock_glis}, which illustrates a single instance of the Monte Carlo runs. The figure shows the points sampled by GLIS and Meta-GLIS over the algorithm's iterations, projected onto three planes as in Fig.~\ref{fig:rosenbrock_autoencoder}. It is evident that in GLIS, which directly optimizes over $x$, the sampled points (blue circles) after 100 iterations almost uniformly cover the search space $\mathcal{X}$. In contrast, in Meta-GLIS, where optimization is performed over a lower-dimensional space, the sampled points (orange circles) remain within the learned manifold by construction, and the solution (orange stars) converges much closer to the optimal value.

Another advantage of Meta-GLIS over GLIS is the time required to optimize the acquisition function: on average, 0.027 s for Meta-GLIS \textit{vs.} 0.149 s for GLIS. This is because the acquisition function in Meta-GLIS is defined over a lower-dimensional space ($n_z=3$) compared to GLIS ($n_x=20$). As for the time required to fit the surrogate model, it is similar for both Meta-GLIS and GLIS (approximately 0.002 s), as it primarily depends on the number of sampled points rather than on the input dimensionality.

\subsection{Automated calibration of MPC hyperparameters}
As another case study, we consider the same problem in  \cite{forgione2020efficient}, which uses GLIS for automatically tuning certain hyperparameters of a model predictive controller.  The system to be controlled is a  cart-pole  described by the differential equations:
\begin{align}
    \label{eq:cart-pole-dyn}
    (M+m)\ddot{p} + mL\ddot{\psi} - mL\dot{\psi}^2\sin{\psi} + b\dot{p} &= F,\\
    L\ddot{\psi} + \ddot{p}\cos{\psi} - g\sin\psi+f_\psi\dot{\psi} &=0,
\end{align}
with $p$, $\psi$ and $F$ being, respectively, the cart position (m), the pendulum angle (rad), and the input force on the cart (N). The force $F$ is regulated by the MPC to track a position reference $p^{\mathrm{ref}}$ (dashed red line in Fig.~\ref{fig:mpc_simulation}), while maintaining the pole in a vertical position (i.e., $\psi \approx 0$) for a pre-defined  duration of the experiment $T^{\mathrm{exp}}$. 
In addition, whenever: $(i)$ the pendulum falls ($|\psi|>\frac{\pi}{6}$), $(ii)$ the cart position exceeds the track length ($|p|> 1.1$ m), or $(iii)$ numerical errors occur in solving the MPC problem, the controller is stopped at a time $T^{\mathrm{stop}} < T^{\mathrm{exp}}$.

For the implementation of the MPC, the continuous-time nonlinear dynamics \eqref{eq:cart-pole-dyn} is discretized  with a sampling time  $\TMPC$, which must be greater than the time required for solving the MPC control law online $\TCALCMPC$.
At each time step $t$, the MPC provides, with a receding-horizon strategy, the  input  minimizing the following  objective function: 
\begin{align} \label{eq:MPC_loss}
J_{\rm MPC} = &  \displaystyle{\sum_{k=0}^{N_p-1}}\Big( (y_{t+k|t} - y^\text{ref}_{t+k})^T Q_y (y_{t+k|t} - y^\text{ref}_{t+k}) \nonumber \\
 &+ (u_{t+k|t} - u^\text{ref}_{t+k})^T Q_u (u_{t+k|t} - u^\text{ref}_{t+k}) \nonumber \\
 & + \Delta u_{t+k|t}^T Q_{\Delta u} \Delta u_{t+k|t}\Big) + Q_\epsilon \eta^2,
\end{align}
subject to input and output constraints (see~\cite{forgione2020efficient} for details). In \eqref{eq:MPC_loss}, $y= [p, \ \psi]$; $u=F$; $N_p$ is the prediction horizon; $u^\text{ref}_{t}$ and $y^\text{ref}_{t}$ are the input and output reference; respectively, at a generic time step $t$; $\eta$ is a slack variable used in \cite{forgione2020efficient} to soften the input and output constraints; and $\Delta u_{t}$ is the  input variation. 

\begin{figure}[h!]
    \centering
    \includegraphics[width=\linewidth]{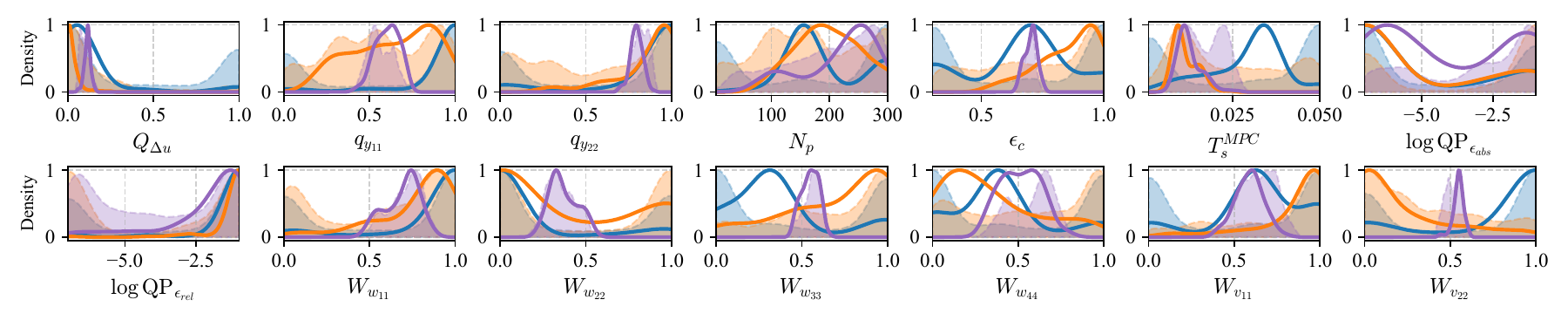}
    \caption{
    Empirical distributions of the optimal solutions over $100$ test system realizations (solid lines) and corresponding $3000$ queries sampled during the optimization (shaded areas) for each MPC hyperparameter: GLIS (blue); Meta-GLIS with nonlinear  (orange) and linear decoder (purple).}
    \label{fig:mpc_distribution_comparison}
\end{figure}

The goal of this case study is tune the following $n_x = 14$ hyperparameters within the intervals visualized in Fig.~\ref{fig:mpc_distribution_comparison}: the positive semidefinite weight matrices of the MPC cost  $Q_y = \bigl( \begin{smallmatrix} q_{y_{11}}&0\\ 0&q_{y_{22}}\end{smallmatrix} \bigr)$, $Q_{\Delta u} \in \mathbb{R}$;  prediction $N_p$  and control horizon $N_u$; MPC sampling time $\TMPC$; log of the absolute and relative tolerances of the QP solver $\log{\text{QP}}_{\epsilon_{abs}}$,  $\log{\text{QP}}_{\epsilon_{rel}}$ used to minimize \eqref{eq:MPC_loss};  process and measurement noises of diagonal covariance matrices $W_w \in \mathbb{R}^{4\times4}$ and $W_v \in \mathbb{R}^{2\times2}$ and used by a Luenberger observer to reconstruct the system state.

The hyperparameter-calibration problem is formalized as the minimization of the following  closed-loop performance index over the entire duration $\Texp$ of the experiment:
\begin{align}
    \label{eq:mpc-cost}
    \tilde{J}^{\mathrm{cl}} = \ln \left ( \int_{t=0}^{\Texp}10 |p^{\mathrm{ref}}(t) - p(t)| +  30 |\phi(t)|\;dt \right )\nonumber\\
  +  \ell\left(\TCALCMPC -  \TMPC\right) + \ell'(\Texp - \Tstop),
\end{align}
where the first term penalizes the tracking error and oscillations of the pendulum, while $\ell$ and $\ell'$ are barrier functions  penalizing, respectively, unfeasible real-time implementation of the control law (i.e., $\TCALCMPC \geq   \TMPC$) and  early stopping of the experiment  (i.e., $\Tstop < \Texp $).

To generate the meta-dataset $\mathcal{D}$,  we sample the parameters $\theta=[M,m,b,f_\varphi,L]^{\top}$ of the cart-pole dynamics equations from the uniform distribution over $\mathcal{U}_{[0.5,2.5]}\times\mathcal{U}_{[0.2,1.0]}\times\mathcal{U}_{[0.1,0.5]}\times\mathcal{U}_{[0.1,0.5]}\times\mathcal{U}_{[0.3,1.5]}$, generating a total of $N=200$ system configurations. In the generation of $\mathcal{D}$, the cost function~\eqref{eq:mpc-cost} is minimized with GLIS over $500$ iterations, leading to $\mathcal{D} = \left\{\theta^{(i)}, \{x^{(i)}_k, f^{(i)}_k\}_{k=1}^{500} \right\}_{i=1}^{200}$  with $200\times 500 = 100\,000$ points. These explored points are used for training an autoencoder with the same architecture as the previous example ($n_z=3$), with $\lambda = 0.5$.

Meta-GLIS is then applied  on $N^{\mathrm{test}} = 100$ test cases, where the cart-pole dynamics are sampled from the same distributions as those used for training. We remark that, in this case study,  we assume a scenario where the meta-dataset used to train the autoencoder is constructed using synthetic data (e.g., simulators with varying settings), while Meta-GLIS should be used for hyperparameter calibration on real (and possibly time-consuming) experiments. 

For a fair comparison, Meta-GLIS is evaluated against  standard GLIS that also leverages information from the meta-dataset $\mathcal{D}$. In particular, we initialize GLIS with  the  optimal  hyperparameter vector associated to the central  parameter 
$\theta_{\mathrm{o}} = [1.5,\allowbreak 0.6,\allowbreak 0.3,\allowbreak 0.3,\allowbreak 0.9]$.
\begin{figure}
    \centering
    \includegraphics[width=\linewidth]{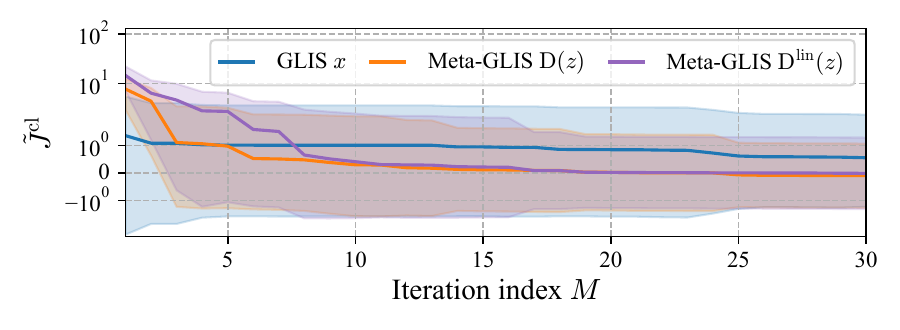}
    \caption{Closed-loop performance index \emph{vs} number of algorithm's iterations. GLIS (blue); Meta-GLIS (organge);  Meta-GLIS with linear decoder (purple). Average (solid lines) and standard deviation (shaded areas) over $100$ system  realizations.}
    \label{fig:mpc_iter_lines_comparison}
\end{figure}
From Fig.~\ref{fig:mpc_iter_lines_comparison}, we can see that Meta-GLIS (orange line) starts with a higher cost with respect to GLIS, but it quickly converges to cost values $\tilde{J}^{\mathrm{cl}}\leq0$ at around iteration 20, where the cart position $p$ has oscillations that are very small around the reference $p^{\mathrm{ref}}$. This can be visualized in the simulations in Fig.~\ref{fig:mpc_simulation},  where the shaded areas associated to GLIS (blue) and Meta-GLIS (orange) represent the standard deviation of the trajectories over  the 100 tests. Indeed,  the orange shaded area is very narrow and almost coincides with the reference signal.
\subsection{Linear Decoder}
\begin{figure}[!tb]
    \centering
    \includegraphics[width=\linewidth]{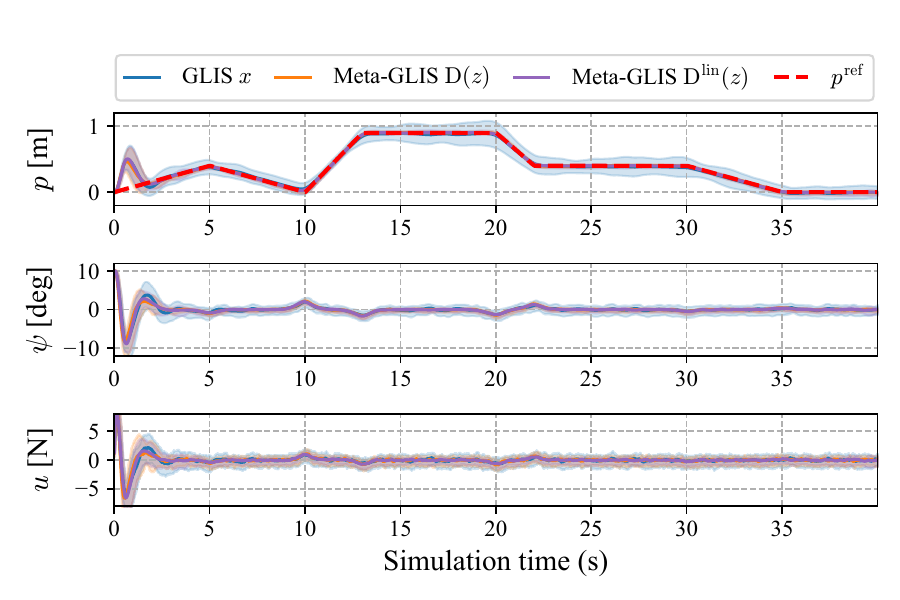}
    \caption{MPC simulation over 100 test systems: achieved performance with hyperparameters calibrated through GLIS (blue); Meta-GLIS (orange) and Meta-GLIS with linear Decoder (purple). Average (solid lines); standard deviations (shaded areas); position reference (dashed red line).}
    \label{fig:mpc_simulation}
\end{figure}

In order to provide an interpretation of the Meta-GLIS algorithm, we modified the architecture of the autoencoder by considering a linear decoder  $\mathrm{D}_{\phi_d}^{\mathrm{lin}}(z)$  of the form $\hat{x} = Az + b$, with $A \in \mathbb{R}^{\nx \times \nz}$ and $b \in \mathbb{R}^{\nx}$. The structure of the encoder and the dimension of the latent variable ($n_z=3$) are not changed with respect to the case discussed in the previous paragraph.   From Figs.~\ref{fig:mpc_iter_lines_comparison} and~\ref{fig:mpc_simulation}, we can see that  Meta-GLIS with linear decoder still outperforms original GLIS, with only a slight performance deterioration with respect to Meta-GLIS with a nonlinear decoder.

\begin{figure}
    \centering
\includegraphics[width=\linewidth, trim=0 0 0 0, clip]{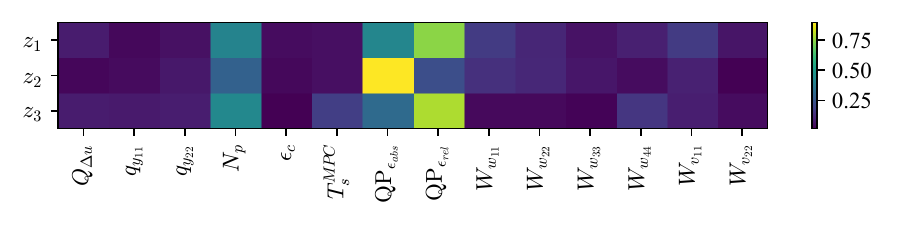}
    \caption{Magnitude of the weights  of the normalized matrix $A^{\top}$ characterizing the linear decoder.}
\label{fig:mpc_lin_magnitude}
\end{figure}

In Fig.~\ref{fig:mpc_lin_magnitude}, we can visualize the magnitude of the trained weights of the matrix $A$, after normalizing its columns to have unit Euclidean norm. In this way, we can see which MPC hyperparameters are more sensitive to a variation of the latent variable $z$. From the figure, we can see that variation of $z$ mainly influences three MPC hyperparameters, namely: $N_p$ and the log-tolerances of the QP solver $\mathrm{QP}_{\epsilon_{abs}}$, $\mathrm{QP}_{\epsilon_{rel}}$. 

The relevance  of each MPC hyperparameter is reflected  
on the empirical distributions of the sampled points during the optimization and of the optimal ones. These distributions are visualized in  Fig.~\ref{fig:mpc_distribution_comparison}. We can observe that, in the $30$ iterations of the algorithm,  GLIS tends to explore  the bounds of the input domains  (shaded blue areas). %, with this characteristic bimodal distribution. 
This behavior is mitigated in Meta-GLIS (shaded orange and purple areas), where the exploration is more concentrated in the region containing the optima (like in $Q_{\Delta u}$, $W_v$), or otherwise more uniform within the search space ($q_y$, $W_{w_{33}}$, $W_{w_{33}}$). Looking at the explored points when using the linear decoder, the exploration remains uniform in the search space for those hyperparameters that are more sensitive to the variation of the latent variable (namely, $N_p$, $\mathrm{QP}_{\epsilon_{abs}}$, $\mathrm{QP}_{\epsilon_{rel}}$).

\section{Conclusions} \label{sec:conclusions}
We presented a method for reducing the dimension of the decision variable space in black-box optimization, thereby decreasing the computational effort required for numerical solution. The core idea of our approach is to learn a low-dimensional embedding within the original decision variable space using an autoencoder trained on precomputed solutions of optimization problems sampled within a given class. These precomputed solutions can be generated, for instance, through simulations with varying parameters and, if necessary, via computationally intensive optimization procedures. Moving from simulation to real, optimization is performed in the embedded space using experiment-driven techniques such as Bayesian Optimization or GLIS. This approach significantly reduces the number of required experiments compared to optimizing directly in the original higher-dimensional space, as we have shown on the Rosenbrock function benchmark and automated calibration of MPC hyperparameters.

Future research will extend the methodology in several directions. On the methodological side, we aim to integrate the approach with preference-based optimization, where objective functions cannot be directly measured but are instead evaluated through qualitative comparisons. On the theoretical side, further work is needed to sharpen the generalization bounds, as well as to extend the analysis to broader meta-learning scenarios such as amortized optimization and system identification in reduced spaces, ultimately providing theoretical tools to assess and enhance the trustworthiness of reduced-space methods.

\bibliographystyle{plain}
\bibliography{references}

\begin{thebibliography}{10}

\bibitem{audet2006mesh}
Charles Audet and John~E Dennis~Jr.
\newblock Mesh adaptive direct search algorithms for constrained optimization.
\newblock {\em SIAM Journal on optimization}, 17(1):188--217, 2006.

\bibitem{Bem20}
A.~Bemporad.
\newblock Global optimization via inverse distance weighting and radial basis functions.
\newblock {\em Computational Optimization and Applications}, 77:571--595, 2020.

\bibitem{forgione2020efficient}
Marco Forgione, Dario Piga, and Alberto Bemporad.
\newblock Efficient calibration of embedded {MPC}.
\newblock {\em IFAC-PapersOnLine}, 53(2):5189--5194, 2020.

\bibitem{gal2010postoptimal}
Tomas Gal.
\newblock {\em Postoptimal Analyses, Parametric Programming, and Related Topics: degeneracy, multicriteria decision making, redundancy}.
\newblock Walter de Gruyter, 2010.

\bibitem{Gut01}
H.M. Gutmann.
\newblock A radial basis function method for global optimization.
\newblock {\em Journal of Global Optimization}, 19:201--227, 2001.

\bibitem{hinton06}
G.~E. Hinton and R.~R. Salakhutdinov.
\newblock Reducing the dimensionality of data with neural networks.
\newblock {\em Science}, 313(5786):504--507, 2006.

\bibitem{jeong2005efficient}
Shinkyu Jeong, Mitsuhiro Murayama, and Kazuomi Yamamoto.
\newblock Efficient optimization design method using kriging model.
\newblock {\em Journal of aircraft}, 42(2):413--420, 2005.

\bibitem{addGP15}
K.~Kandasamy, J.~Schneider, and B.~P\'{o}czos.
\newblock High dimensional bayesian optimisation and bandits via additive models.
\newblock In {\em Proc. of the 32nd International Conference on Machine Learning}, page 295–304. JMLR.org, 2015.

\bibitem{letham20}
B.~Letham, R.~Calandra, A.~Rai, and E.~Bakshy.
\newblock Re-examining linear embeddings for high-dimensional bayesian optimization.
\newblock In {\em Advances in Neural Information Processing Systems}, volume~33, pages 1546--1558. Curran Associates, Inc., 2020.

\bibitem{massart1990tight}
Pascal Massart.
\newblock The tight constant in the {Dvoretzky--Kiefer--Wolfowitz} inequality.
\newblock {\em The Annals of Probability}, 18(3):1269--1283, 1990.

\bibitem{MBC79}
M.D. McKay, R.J. Beckman, and W.J. Conover.
\newblock Comparison of three methods for selecting values of input variables in the analysis of output from a computer code.
\newblock {\em Technometrics}, 21(2):239--245, 1979.

\bibitem{rolland18additive}
P.~Rolland, J.~Scarlett, I.~Bogunovic, and V.~Cevher.
\newblock High-dimensional bayesian optimization via additive models with overlapping groups.
\newblock In {\em Proc. of the 21st Intl. Conf. on Artificial Intelligence and Statistics}, volume~84 of {\em Proc. of Machine Learning Research}, pages 298--307. PMLR, 09--11 Apr 2018.

\bibitem{Ros60}
H.~H. Rosenbrock.
\newblock An automatic method for finding the greatest or least value of a function.
\newblock {\em The Computer Journal}, 3(3):175--184, 01 1960.

\bibitem{RoFoPi20}
Loris Roveda, Marco Forgione, and Dario Piga.
\newblock Robot control parameters auto-tuning in trajectory tracking applications.
\newblock {\em Control Engineering Practice}, 101, 2020.

\bibitem{sabug2022smgo}
Lorenzo Sabug~Jr, Fredy Ruiz, and Lorenzo Fagiano.
\newblock {SMGO-$\Delta$:} balancing caution and reward in global optimization with black-box constraints.
\newblock {\em Information Sciences}, 605:15--42, 2022.

\bibitem{schmidhuber1987evolutionary}
J{\"u}rgen Schmidhuber.
\newblock {\em Evolutionary principles in self-referential learning, or on learning how to learn: the meta-meta-... hook}.
\newblock PhD thesis, Technische Universit{\"a}t M{\"u}nchen, 1987.

\bibitem{Shahriari16}
B.~Shahriari, K.~Swersky, Z.~Wang, R.~P. Adams, and N.~de~Freitas.
\newblock Taking the human out of the loop: A review of bayesian optimization.
\newblock {\em Proceedings of the IEEE}, 104(1):148--175, 2016.

\bibitem{storn1997differential}
Rainer Storn and Kenneth Price.
\newblock Differential evolution--a simple and efficient heuristic for global optimization over continuous spaces.
\newblock {\em Journal of global optimization}, 11:341--359, 1997.

\bibitem{REMBO16}
Z.~Wang, F.~Hutter, M.~Zoghi, D.~Matheson, and N.~De~Freitas.
\newblock Bayesian optimization in a billion dimensions via random embeddings.
\newblock {\em J. Artif. Int. Res.}, 55(1):361–387, 2016.

\bibitem{ziomek23a}
J.~K. Ziomek and H.~Bou~Ammar.
\newblock Are random decompositions all we need in high dimensional {B}ayesian optimisation?
\newblock In {\em Proc. of the 40th International Conference on Machine Learning}, volume 202 of {\em Proceedings of Machine Learning Research}, pages 43347--43368. PMLR, 2023.

\end{thebibliography}

%\addtolength{\textheight}{-12cm} 

\section*{Appendix}

In this appendix we provide the proof of Theorem \ref{theorem:gap}. First, a set of useful result is provided.
\begin{property}[i.i.d.\ structure of the performance gap]\label{prop:iid-Psi}
Let $(\theta^{(i)},\xi^{(i)},\omega^{(i)})$, $i=1,\ldots,m$, be i.i.d.\ draws from the product 
$p(\theta)\otimes p(\xi)\otimes p(\omega)$. 
Define the performance gaps
\begin{align}
\Psi^{(i)} \;:=\; \Psi\big(\theta^{(i)},\xi^{(i)},\omega^{(i)}\big), 
\qquad i=1,\dots,m,
\end{align}
as in \eqref{eqn_perc_gap}.
Then $\{\Psi^{(i)}\}_{i=1}^m$ are i.i.d.\ real-valued random variables. %\hfill $\blacksquare$
\end{property}

\paragraph*{Proof of Property \ref{prop:iid-Psi}}
By hypothesis, $(\theta^{(i)},\xi^{(i)},\omega^{(i)})$ are i.i.d.\ samples drawn from the product distribution 
$p(\theta)\otimes p(\xi)\otimes p(\omega)$. 
For each triple $(\theta,\xi,\omega)$, the performance gap $\Psi(\theta,\xi,\omega)$ is obtained by applying the 
same deterministic transformation, consisting of the optimization algorithms $\mathcal A_x$ and $\mathcal A_z$, the 
decoder $\emb$, and the objective function $f$. 
Since applying a fixed deterministic transformation to i.i.d.\ inputs preserves both independence and identical distribution, 
the resulting values 
\begin{align}
\Psi^{(i)} = \Psi(\theta^{(i)},\xi^{(i)},\omega^{(i)}), \qquad i=1,\dots,m,
\end{align}
are themselves i.i.d.\ real-valued random variables. \hfill $\blacksquare$

The following lemma by Dvoretzky–Kiefer–Wolfowitz (DKW), and refined by Massart in~\cite{massart1990tight}
 establishes that the empirical cumulative distribution function  
$\hat F_m$ of $m$ i.i.d.\ samples uniformly approximates the true cumulative distribution function   $F$ of the underlying random variable $\Psi$. 
Specifically, with probability at least $1-\delta$, the maximum deviation 
$\sup_t |\hat F_m(t) - F(t)|$ is bounded by $\varepsilon_m = \sqrt{\log(2/\delta)/(2m)}$. 
This result provides a non-asymptotic, distribution-free guarantee on the accuracy of the empirical CDF.

\begin{lemma}[DKW--Massart inequality~\cite{massart1990tight}]\label{lem:DKW}
Let $\Psi$ be a real-valued random variable with \emph{cumulative density function} (CDF) $F$ and let
$\hat {F}_m$ be the empirical CDF of  i.i.d. samples $\Psi^{(1)},\ldots,\Psi^{(m)}$ defined as:
\begin{align} \label{eqn:ECDF}
    \hat {F}_m(t) = \frac{1}{m}\sum_{i=1}^m\mathbf{1}[\Psi^{(i)} \leq t], \ \ t \in \mathbb{R},
\end{align}
with $\mathbf{1}$ denoting the indicator function.

Then, for any $\delta\in(0,1)$, with
\begin{align}
\varepsilon_m \ :=\ \sqrt{ \frac{\log(2/\delta)}{2m} },
\end{align}

it holds that
\begin{align}
\mathbb P_S\!\left( \sup_{t\in\mathbb R}\, \big| \hat{F}_m(t) - F(t) \big| \ \le \ \varepsilon_m \right) \ \ge \ 1-\delta,
\end{align}
where  $S := \{\Psi^{(i)}\}_{i=1}^m$ is the  set of i.i.d.\ samples of $\Psi$. 
%\hfill $\blacksquare$

\end{lemma}

In the following section, we provide the proof of Theorem \ref{theorem:gap}.

\paragraph*{Proof of Theorem \ref{theorem:gap}} 
We proceed in two steps.

\textbf{Step 1.} 
The goal of this step is to show that the true quantile 
$Q_{1-\alpha}:=\inf\{t: F(t)\ge 1-\alpha\}$ (for any $\alpha \in (0,1)$) 
can be upper bounded, with high probability (i.e., with probability at least $1-\delta$, for arbitrary $\delta \in (0, 1)$), by a suitable order statistic of the validation samples 
$S=\{\Psi^{(i)}:=\Psi(\theta^{(i)},\xi^{(i)},\omega^{(i)})\}_{i=1}^m$, with 
%where 
\begin{align}
 (\theta^{(i)},\xi^{(i)},\omega^{(i)}) \sim p(\theta)\otimes p(\xi)\otimes p(\omega).
\end{align}

Let $t^\star := \Psi_{(k^\star)}$ be the $k^\star$-th order statistic of the validation set. 
By construction, at least $k^\star$ samples satisfy $\Psi^{(i)} \le t^\star$, hence by definition of the empirical CDF in eq. \eqref{eqn:ECDF}, we have:
\begin{align}
    \hat{F}_m(t^\star)\ \ge\ \frac{k^\star}{m},
\end{align}
where the inequality (and not equality) holds because multiple samples $\Psi^{(i)}$ may coincide with $t^\star$.

Define the event
\begin{align}
\mathcal E := \Big\{\sup_{t\in\mathbb R}|\hat{F}_m(t)-F(t)|\le \varepsilon_m\Big\}.
\end{align}
On $\mathcal E$, we have
\begin{align} \label{eqn:Fineq}
F(t^\star)\ \ge\ \hat{F}_m(t^\star)-\varepsilon_m\ \ge\ \frac{k^\star}{m}-\varepsilon_m\ \geq \ 1-\alpha, 
\end{align}
where the last inequality follows from the definition of $k^\star$. 

By  definition of the true quantile, inequality \eqref{eqn:Fineq} implies
\begin{align}
Q_{1-\alpha}\ \le\ t^\star \ =\ \Psi_{(k^\star)}.
\end{align}

Let us set $\varepsilon_m := \sqrt{ \frac{\log(2/\delta)}{2m} }$. Since $\{\Psi^{(i)}\}_{i=1}^m$ are i.i.d.\   random variables (Property \ref{prop:iid-Psi}),  the DKW–Massart inequality (Lemma~\ref{lem:DKW}) can be applied. This implies that for any $\delta\in(0,1)$, the event $\mathcal E$ holds with probability at least $1-\delta$ over the draw of the validation set $S$. 
Thus, 
\begin{align}
\mathbb P_S\!\left( Q_{1-\alpha}\ \le\ \Psi_{(k^\star)} \right) \ \ge\ 1-\delta.
\end{align}

\textbf{Step 2.} 
We now leverage the quantile bound from Step~1 to control the performance on a new draw 
$(\theta^{\rm new},\xi^{\rm new},\omega^{\rm new}) \sim p(\theta)\otimes p(\xi)\otimes p(\omega)$. 
By definition of the $(1-\alpha)$-quantile,
\begin{align}
\mathbb P_\Psi\!\big(\Psi^{\rm new} \le Q_{1-\alpha}\big) = 1-\alpha.
\end{align}

If $\mathcal E$ holds, then $Q_{1-\alpha}\le \Psi_{(k^\star)}$, and hence
\begin{align}
  \{\Psi^{\rm new} \le \Psi_{(k^\star)}\} \ \supseteq \ \{\Psi^{\rm new} \le Q_{1-\alpha}\} .
\end{align}
Therefore,
\begin{align}
\mathbb P_\Psi\!\left(\Psi^{\rm new} \le \Psi_{(k^\star)} \,\Big|\, S\right)  
&\;\ge\; \mathbb P_\Psi\!\big(\Psi^{\rm new} \le Q_{1-\alpha}\big)  \nonumber \\
&=\; 1-\alpha.
\end{align}

Taking the outer probability over the validation samples $S$, we obtain
\begin{align}
\mathbb P_S\!\Big(
   \mathbb P_\Psi\!\left(\Psi^{\rm new} \le \Psi_{(k^\star)} \,\Big|\, S\right) \ge 1-\alpha
\Big)\;\ge\;1-\delta,
\end{align}
which completes the proof. \hfill $\blacksquare$

\end{document}